# Enhancing Chemistry Learning with ChatGPT and Bing Chat as Agents-to-Think-With: A Comparative Case Study

Renato P. dos Santos

CIAGE – Centre for Generative Artificial Intelligence in Cognition and Education

ULBRA – Lutheran University of Brazil

E-mail: RenatoPSantos@ulbra.edu.br

**ABSTRACT**

This study explores the potential of Generative AI chatbots (GenAIbots) such as ChatGPT and Bing Chat, in Chemistry education, within a constructionist theoretical framework. A single-case study methodology was used to analyse extensive interaction logs between students and both AI systems in simulated Chemistry learning experiences. The results highlight the ability of ChatGPT and Bing Chat to act as 'agents-to-think-with', fostering critical thinking, problem-solving, concept comprehension, creativity, and personalised learning experiences. By employing a Socratic-like questioning approach, GenAIbots nurture students' curiosity and promote active learning. The study emphasises the significance of prompt crafting, a technique to elicit desired responses from GenAIbots, fostering iterative reflections and interactions. It underlines the need for comprehensive educator training to effectively integrate these tools into classrooms. The study concludes that while ChatGPT and Bing Chat as agents-to-think-with offer promising avenues to revolutionise STEM education through a constructionist lens, fostering a more interactive, inclusive learning environment and promoting deeper comprehension and critical thinking in students across diverse Chemistry topics, ChatGPT consistently outperformed Bing Chat, providing more comprehensive, detailed, and accurate responses and skillfully addressing nuances and context.

**CONTRIBUTION TO THE LITERATURE**

- This study fills a gap in current literature by investigating ChatGPT and Bing Chat's roles in enhancing Chemistry learning experiences.

- The application of ChatGPT and Bing Chat as agents-to-think-with, resources that foster reflective and critical thinking, problem-solving skills, and concept comprehension, is examined within the context of Chemistry education, offering a novel perspective to the field.
- It is shown that, notably, ChatGPT consistently outperformed Bing Chat, providing more comprehensive, detailed, and accurate responses and skillfully addressing nuances and context.
- Logs of students' extensive interactions with ChatGPT and Bing Chat are presented and analysed, providing valuable data on AI-assisted Chemistry learning experiences.

**INTRODUCTION**

Chemistry is a vital subject in Science Education. It serves as a foundation for comprehending the natural world by elucidating matter's composition, properties, behaviour, and transformations, thus playing a crucial role in our daily lives (Dunlop et al., 2020). However, Chemistry is also a complex and dynamic field that necessitates a profound grasp of fundamental concepts and principles. Nevertheless, it is difficult for students to associate real-life circumstances with abstract chemistry concepts (Dewi et al., 2021).

Chemistry Education research is mostly about making students understand chemistry concepts better and creating meaningful Chemistry learning. Thus, it is typically focused on learning strategies and media to help students understand concepts and other learning factors, including instruction and evaluation (Dewi et al., 2021).

Despite Chemistry Education's importance, research reveals that numerous students struggle with learning Chemistry due to factors such as its complexity, abstract nature, misconceptions, lack of prior knowledge, and suboptimal teaching methods (Dunlop et al., 2020; Timilsena et al., 2022; Tümay, 2016), resulting in diminished interest and engagement in the subject (Dewi et al., 2021). In recent years, the high failure rates and low achievement levels among high school Chemistry students have become a significant concern in Chemistry Education.

To address this long-standing issue, Science Education researchers have been investigating students' challenges in learning Chemistry and exploring effective ways to address them (Permatasari et al., 2022; Timilsena et al., 2022; Tümay, 2016).

Given these challenges, Generative Artificial Intelligence-powered chatbots (GenAIbots), such as OpenAI's ChatGPT[1] and Microsoft's Bing Chat[2], hold the potential to help students overcome these difficulties. These chatbots offer a more interactive and personalised learning experience, allowing students to delve into complex concepts through engaging dialogue, thereby fostering a deeper understanding of Chemistry. As "agents-to-think-with," agents that foster reflective and critical thinking and concept comprehension in enhancing Education, these GenAIbots enable learners to actively construct their knowledge while receiving real-time feedback and guidance. Furthermore, the conversational nature of chatbots promotes collaboration and critical thinking, both of which are crucial skills for mastering Chemistry.

By incorporating GenAIbots into the Chemistry learning process, educators can create a more dynamic and inclusive learning environment, ultimately addressing the prevalent issues of high failure rates, low achievement levels, and disengagement.

**LITERATURE REVIEW**

Several studies have investigated the reasons behind students' difficulties in learning Chemistry. Some researchers have attributed these difficulties to the abstract nature of chemical concepts, which are often challenging for students to visualise and understand. Others have pointed out that the language used in Chemistry textbooks and lectures can be complex and technical, making it difficult for students to comprehend.

One such study is the research conducted by Timilsena et al. (2022) on teachers' and students' experiences with Chemistry learning difficulties. The study aimed to explore high school students' challenges when learning Chemistry and identify potential solutions for improving Chemistry Education. The researchers used a qualitative research design to collect data from three public schools and their six Science teachers. They also collected 21 students' in-depth interviews, open-ended questionnaires, focus group discussions, and classroom observations. The study's main findings on students' experiences in learning Chemistry were that students find it hard to understand the idea of a chemical reaction due to its abstract nature. The study identified several factors that contribute to learning difficulties, including inadequate teaching materials or methods, unclear curriculum scope and order, and the complexity of the subject. The study also found that most of the teacher respondents who expressed difficulty in Chemistry learning related to students' failure to understand the concepts of Chemistry. Overall, the study highlights the need for effective teaching strategies and tools for improving student understanding of Chemistry concepts.

---

[1] https://openai.com/
[2] https://www.bing.com/new

The paper by Dunlop et al. (2020) aims to explore the difficulties that undergraduate Chemistry students face in their studies and propose potential solutions to address these challenges. The lack of philosophical dialogue in Chemistry Education is a missing component limiting students' ability to achieve happiness and well-being by avoiding discomfort and difficulty. The paper draws on ethnographic observations, questionnaires, and in-depth semi-structured individual interviews to present a model for introducing philosophical dialogue into Higher Education. Furthermore, the impact of Talking Chemistry, an extracurricular space created for undergraduate Chemistry students to build capabilities through philosophical dialogue about Chemistry, is examined. The study suggests that this approach can challenge students to think about their subject in new ways and contribute to their overall well-being.

Tümay (2016) argues that many students struggle with understanding fundamental concepts in Chemistry due to various factors such as misconceptions, lack of prior knowledge, and inadequate teaching methods. In particular, students have difficulty understanding the emergent properties of macroscopic aggregates and often attribute macroscopic properties and changes shown by bulk matter to microscopic entities (atoms, molecules, etc.). Furthermore, the paper highlights the importance of identifying students' misconceptions and learning difficulties in understanding scientific models in Chemistry Education research. It also emphasises the need for educators to consider the role of emergence in their teaching practices to better support students' learning of Chemistry.

Dewi et al. paper (2021) focuses on the urgency of digital literacy for Generation Z students in Chemistry learning. The paper highlights students' difficulties in learning basic Chemistry concepts and how 21st-century skills, such as digital literacy, are increasingly crucial for success in all fields. The research presented here emphasises developing critical thinking skills through reading, writing, understanding, evaluating, communicating, and using information in different formats. The paper identifies three chemical representations - macro, micro, and symbolic - that need to be addressed by Chemistry teachers to ensure that students understand Chemistry concepts fully. Despite this need for a comprehensive approach to teaching Chemistry concepts and digital literacy skills, many students still rely on memorisation rather than understanding concepts. The paper explores how digital literacy can help students overcome these difficulties and provides solutions to improve Chemistry Education quality. It also examines how educators can incorporate digital technology as learning tools and methods for Generation Z students while discussing potential challenges in implementing digital literacy in Chemistry Education and suggesting ways to address them.

Permatasari et al. (2022) conducted a literature review that comprehensively synthesises studies on Chemistry learning using multiple representations. The article addresses the definition of multiple representations, their effect on Chemistry learning outcomes, and how to implement them in Chemistry learning models or strategies. Based on the results of the reviewed articles, it is essential to involve multiple representations in chemical learning to increase students' learning outcomes. Using representations can reduce learners' assumptions about Chemistry as complex and make abstract concepts more real and easier to understand. This review provides valuable insights for teachers and researchers in Chemistry Education, including trends and patterns in Chemistry learning using multiple representations and directions for further investigation.

Resources offered by ChatGPT and Bing Chat can help address these challenges by simplifying complex language, making concepts more accessible, and incorporating multiple representations to facilitate understanding. These GenAIbots can provide interactive simulations and visualisations to make abstract concepts concrete while offering a platform for philosophical dialogues and resource sharing. By fostering innovative teaching strategies and methods, GenAIbots can significantly enhance Chemistry Education, addressing learning difficulties and fostering a deeper understanding of complex concepts.

**THEORETICAL FRAMEWORK**

The idea of "agents-to-think-with" forms a central part of our discourse, an extension of the concept of "objects-to-think-with," initially put forth by Seymour Papert in his classic book "Mindstorms." He described these as "any tool or resource that learners can use to explore and develop their understanding of a particular concept or domain" (1980, p. 11). In Papert's words, "thinking about thinking turns the child into an epistemologist, an experience not even shared by most adults." (Papert, 1980, p. 19).

To put it differently, the concept of objects-to-think-with is a way to describe how specific tools or objects can aid individuals in their learning and cognitive development by engaging the user in dynamic conversations, providing information, asking questions, fostering reflective and critical thinking, improving problem-solving abilities, and deepening the understanding of concepts. In this sense, a wide array of things can serve as objects-to-think-with, such as playing cards, photographs, school maps, sticky notes, Lego blocks, cameras, smartphones, simulation software.

Taylor et al. (2022) underscore the significance of the autonomous role of chatbots and other digital tools in educational research. The authors argue that by attending to the agency of objects, researchers can gain new insights into how learning occurs and, educational practices can be improved and how these tools can be used to enhance learning outcomes.

However, through the lens of the Actor-Network Theory (ANT) (Crawford, 2020), developed by Michel Callon, Madeleine Akrich, Bruno Latour, and John Law, these chatbots must be seen as active participants in the networks that produce knowledge and action, as they interact with human users, process inputs, generate outputs, and even adapt to new information, acting as non-human actors within the network.

Latour's works often deal with the broader concept of hybridity between humans and non-human actors, including technological artifacts. His book "We Have Never Been Modern" (Latour, 1991) discusses the concept of "quasi-objects" or "hybrids" that blur the line between nature and society, indicating a perspective that could be applicable to AI. In "Reassembling the Social: An Introduction to Actor-Network Theory" (Latour, 2005), he further elaborates on the role of non-human actors in social networks.

Latour's idea of hybridization between AI and humans suggests a further blurring of the boundaries between humans and non-humans. This concept posits that thinking and action are the results of interactions within networks that include both human and non-human actors. In this perspective, AI chatbots like ChatGPT or Bing Chat are part of this hybrid network of cognition and action.

As such, AI chatbots can be seen as contributing to our collective thinking in ways that go beyond their role as mere tools. They can generate novel responses, offer new perspectives, and help us see things in different ways. This aligns with Latour's critique of the division between thinking humans and non-thinking non-humans, as it recognizes the active role these non-human entities can play in our collective cognition and action.

We, therefore, propose to extend the concept of "objects-to-think-with," as proposed by Seymour Papert, to "agents-to-think-with" for these Generative Artificial Intelligence-powered chatbots, such as ChatGPT and Bing Chat, as they can be seen as a form of hybridization between non-human AI actors and humans, as suggested in Actor-Network Theory (ANT).

This view aligns with posthumanist theories that stress the intricate interweaving of humans and non-human entities (Haraway, 1990). Accordingly, Swan (2015) envisions a future realm where multiple forms of intelligence coexist productively, including unmodified humans, augmented humans, and various types of artificial intelligence.

People are drawn to those chatbots' ability to provoke and make self-reflection enjoyable. To paraphrase Turkle, chatbots are 'metacognitive machines' because they influence how we think about our own thinking (1984, p. 17).

In the last few months, GenAIbots, such as ChatGPT and Bing Chat, have become increasingly popular as educational tools in Chemistry learning (Baidoo-Anu & Owusu Ansah, 2023; Taylor et al., 2022). These tools have the potential of enhancing learning by providing personalised experiences for students.

We understand that GenAIbots can act as influential agents-to-think-with in Chemistry education, promoting profound learning and comprehension. They provide interactivity, personalisation, multidisciplinary knowledge, critical thinking, problem-solving, creativity, collaboration, and communication (Okonkwo & Ade-Ibijola, 2021). Users participate in dynamic conversations, receive real-time feedback, and enrich reflective-thinking learning experiences.

GenAIbots cater to individual needs, possibly making learning more engaging and relevant while covering a broad range of subjects, allowing users to explore various disciplines. Interaction with GenAIbots fosters cognitive development and in-depth understanding, stimulate users' creativity, generates ideas, and inspires projects (Baidoo-Anu & Owusu Ansah, 2023). GenAIbots facilitate discussions and collaboration in group settings leading to a more profound grasp of the subject matter.

Incorporating GenAIbots into collaborative learning environments fosters dynamic, inclusive spaces where students can exchange ideas, challenge assumptions, and learn collaboratively (Okonkwo & Ade-Ibijola, 2021). GenAIbots' ability to generate practice problems, quizzes, explanations, and examples of Chemistry concepts promotes active learning and enables deeper understanding.

The effectiveness of GenAIbots for generating natural language responses in educational contexts has been well-established (Okonkwo & Ade-Ibijola, 2021; Sabzalieva & Valentini, 2023). ChatGPT, a cutting-edge natural language generation model, excels at producing human-like responses suitable for chatbots (Franciscu, 2023). As a result, GenAIbots offer potential for diverse applications within CHEMISTRY learning, including engagement with hypothetical scenarios, experimentation with problem-solving strategies, and guidance on specific concepts.

Mollick and Mollick (2022) suggest that GenAIbots, such as ChatGPT and Bing Chat, can address three critical educational challenges in classrooms: facilitating the transfer, correcting students' overestimation of their understanding of specific concepts, and nurturing critical evaluation of explanations.

Nonetheless, it is essential to exercise caution when utilising GenAIbots. Firstly, they should not be treated like Google, as relying on them for searches may yield inaccurate results due to their tendency to 'hallucinate,' i.e., fabricate facts and references (OpenAI, 2023, p. 6). Additionally, they are not calculators and may provide incorrect results for mathematical computations. GenAIbots are also distinct from Alexa, making them unsuitable for weather forecasts or jokes. Furthermore, seeking justifications for their incorrect responses may lead to seemingly logical but ultimately false explanations.

Frieder et al. (2023) examined ChatGPT's mathematical proficiency, determining that despite numerous positive media accounts, its capabilities fall considerably short of an average mathematics graduate student. ChatGPT often comprehends the question but fails to deliver accurate solutions. MacIsaac (2023) recounts unproductive interactions with ChatGPT while attempting to solve physics homework problems. Moreover, Borji

(2023) identifies and analyses eleven categories of ChatGPT shortcomings, encompassing reasoning, factual errors, mathematics, coding, and bias.

Consequently, we suggest that employing GenAIbots as agents-to-think-with provides students with exclusive chances to interact with intricate ideas, thereby leading to the emergence of fresh insights and viewpoints. Conversing with these GenAIbots allows students to pose questions, obtain immediate feedback, and delve into diverse perspectives.

On the other hand, despite obtaining the desired answer, the students must exercise their critical thinking skills as GPT-4 has limitations, being able to "hallucinate", i.e., create content that does not make sense or is not truthful compared to specific sources (OpenAI, 2023, p. 6).

## MATERIALS AND METHODS

This section describes the research methodology utilised for this case study. This investigation aims to determine how effectively two complex AI language models, ChatGPT and Bing Chat, can enhance the learning of Chemistry when used as an agent-to-think-with. Our analysis compared their performance as they responded to a series of chemistry-related questions. To ensure a rigorous examination, we have employed a qualitative research design, focusing on an in-depth exploration of participants' experiences and perceptions.

To comprehensively understand students' experiences with GenAIbots as agents-to-think-with in Chemistry learning, this study applied an exploratory holistic single case study research design featuring two distinct units, ChatGPT and Bing Chat. In line with the principles set forth by Yin (2011), this methodology allows for comparing and analysing these units, highlighting the differences in their abilities to respond to the student's questions. The chosen approach provides valuable insights into the dynamics and intricacies of student engagement with GenAIbots, including specific interactions with ChatGPT and Bing Chat.

The goal with GenAIbots is not to pose a series of disconnected questions for which they provide ready-made answers. Mollick (2023) recommends engaging in a dialogue with the AI, asking it questions, experimenting with different prompt formats, giving feedback, and challenging it to improve. This dynamic may resemble a philosophical dialogue involving a sequence of interactions where each response triggers further user reflection, sparks new questions, and cultivates a more profound comprehension.

This process of giving plain language instructions to a language model like GPT is called *prompting* (Mishra et al., 2023). It is a vital part of the process of successive reflections and interactions with the GenAIbots that characterise its use as agents-to-think-with.

Wei et al. (2022) and Kojima et al. (2022) have shown that chain-of-thought (CoT) prompting, sequences of eight or more short sentences describing intermediate reasoning steps towards a final answer, is an effective method for enhancing reasoning in large language models like ChatGPT and Bing Chat. CoT prompting allows these models to perform complex multi-step reasoning tasks and generate more accurate responses, making them effective zero-shot reasoners. Additionally, Fu et al. (2023) suggest that selection based on reasoning complexity is an intuitive and efficient way to improve the performance of these systems in real-world applications.

It is worth noticing that Dunlop et al. argue that philosophical dialogue is a missing component of Chemistry Education and present a model for introducing it into Higher Education (2020). In this context, GenAIbots, such as ChatGPT and Bing Chat, can act as agents-to-think-with, enabling students to engage in philosophical dialogues related to Chemistry. Their conversational nature encourages learners to question and reflect on the subject's underlying principles, assumptions, and ethical implications, fostering critical thinking and a deeper understanding of the discipline. By facilitating such meaningful discussions, GenAIbots can help bridge the gap between theoretical knowledge and its practical and philosophical aspects, providing a more holistic and well-rounded educational experience for students in Chemistry.

**Participants**

Using ChatGPT or Bing Chat as an agent-to-think-with is relatively new and innovative. Finding teachers willing and prepared to incorporate it into their classrooms for experimental purposes has proven challenging. Consequently, the investigation did not involve any real students. Instead, the researcher acting as a STEM teacher at a Brazilian university, acted as a student in the following interaction sessions. As a STEM teacher, they provided valuable insights from an instructional perspective. However, to preserve privacy, they will be identified as P1.

**Procedure**

The participant conducted two sessions where they simulated Chemistry students' learning experiences by interacting one with ChatGPT and another with Bing Chat.

Based on the previous discussion on prompting, our approach was influenced by the idea that the most effective way to learn with AI is by treating it as a partner and a teacher (Mollick, 2023). Various prompts were experimented with to optimise our interactions, providing feedback to the GenAIbots and challenging them to improve before settling on the prompt used, recognising it may not be the best choice.

**Instruments**

The study employed two primary instruments for data collection:

- GenAIbot interaction logs: Logs from participants' interactions with GenAIbots were collected, recording their sequences of interactions with prompts and responses, allowing for later analyses of the content, quality, and nature of the exchanges.
- Reflective journals: Participants maintained reflective journals throughout the study, documenting their experiences, challenges, and learning outcomes when using a GenAIbot in their Chemistry learning activities.

**Data Analysis**

Initially, the researcher familiarised themselves with the data through repeated readings of the GenAIbots interaction logs and reflective journals. Following this, those materials were analysed to identify recurrent themes and patterns, thereby elucidating the perceived impact of GenAIbots as an agent-to-think-with on Chemistry learning.

**RESULTS AND ANALYSES**

This section contains excerpts from the interaction logs with the GenAIbots for comparative purposes (the full versions of the logs can be accessed on figshare[3]). Since ChatGPT and Bing Chat provide unique responses, it is impossible to follow the exact sequence of questions for comparison. Moreover, analyses of the findings from the sessions where participants simulated Chemistry students' learning experiences by engaging with the GenAIbots.

---
[3] https://doi.org/10.6084/m9.figshare.22798418.v1

This study aimed to investigate the effectiveness of ChatGPT and Bing Chat in enhancing Chemistry learning. Empathy was incorporated into the research methodology to enable the researchers to act like students and understand their perspectives.

**1st session – Understanding Chemistry**
GenAIbot: ChatGPT:
Model: GPT-4
Release date: May-03
Participant: P2
Usage time: 2023-May-04.

**2nd session – Understanding Chemistry**
GenAIbot: Bing Chat
Model: not available
Release date: not available
Participant: P2
Usage time: 2023-May-10.

**Prompts and generated texts** (translated from Portuguese for convenience)

*P2: I will start a dialogue where I will assume the role of a student, engaging with you,*
ChatGPT*: as an agent-to-think-with, with the primary objective of deepening and deliberating on Chemistry concepts. As an agent-to-think-with, I kindly request your support in providing engaging and personalised responses to student questions. It is critical that our interaction adopt a Socratic approach, encouraging the student to reflect on their existing knowledge, encouraging critical thinking, including examples and analogies where possible, and helping to refine their understanding rather than offering ready-made answers. Are you ready to embark on this collaborative and thought-provoking journey?*

**ChatGPT:** *Yes, I'm ready to embark on this collaborative and thought-provoking journey with you. Let's explore Chemistry concepts together, using a Socratic approach and focusing on developing your critical thinking. Please feel free to ask your questions and share your thoughts. I'm here to help you deepen your understanding.*

**Bing Chat:** *Hello! I understood that you would like to have a Socratic dialogue with me to help students reflect on their existing knowledge in Chemistry. I'm ready to embark on this collaborative and thought-provoking journey! Let's start?* 😊

*P2: I have difficulty with this business of chemical reactions and balance*

**ChatGPT:** *I understand that balancing chemical reactions can be a challenging topic. Let's tackle it step by step.*

*A chemical reaction is a transformation where substances are converted into other substances, involving the breaking and formation of chemical bonds. The balance of a chemical reaction is a process by which the amounts of atoms of each element involved in the reaction are equalized. Think of it as a scale: on one side we have the reactants (the substances that react) and on the other side we have the products (the substances formed). For the balance to be balanced, the number of atoms of each element must be equal on both sides.*

*Now, consider an analogy: imagine that you have several boxes of different sizes and you need to stack them evenly on two sides of a shelf. This shelf represents the chemical equation. To balance the shelf, you adjust the number of boxes on each side until the shelf is balanced. The balance of a chemical reaction follows the same principle.*

*Can you give me an example of a chemical reaction that you would like to balance or delve a little deeper into the concepts involved before moving on?*

**Bing Chat:** *I understand that you have difficulty balancing chemical reactions. Balancing chemical equations is one of the most important skills in chemistry. When writing an equation for a chemical reaction, both sides of the equation must balance - you need the same number of each type of element on both sides [1].*

*Balancing can be accomplished in several ways. The main methods are: trial, redox and algebraic [2]. The algebraic method is the most common and involves assigning coefficients to chemical species so that the total number of atoms of each element is equal on both sides of equation [3].*

*Would you like me to provide a practical example to help you better understand how to balance a chemical reaction?*

Overall, both responses are helpful, but ChatGPT's explanation might be more engaging and accessible due to its analogy and step-by-step approach. It also asks the student for their preference, which can be helpful in tailoring the explanation to the student's needs. Bing Chat, on the other hand, provides a more technical explanation and offers to provide a practical example, which can be useful for students who prefer a hands-on approach.

**P2:** Yeah, but that sounds like my teacher talking. I don't even understand this chemical reaction business. What's the use of an example of a business I don't understand?

**ChatGPT:** *I apologize if the previous explanation was not clear. Let's try a different approach. A chemical reaction is like a "change" that occurs when different substances interact with each other. For example, when you mix vinegar and baking soda, you see bubbling and fizzing –*

*that's a chemical reaction taking place. The initial substances (vinegar and sodium bicarbonate) are being transformed into new substances (carbon dioxide, water and sodium acetate).*

*Now, imagine you are a cook and you have some ingredients (reagents) to make a dish (products). Each ingredient has a certain amount to be used in the recipe. Balancing the chemical reaction is similar to following a recipe, ensuring you are using the correct amount of each ingredient.*

*So I'll give you an example of a chemical reaction that you might have heard of: the combustion of methane (CH4). Methane reacts with oxygen (O2) to form carbon dioxide (CO2) and water (H2O). The unbalanced equation is:*

CH4 + O2 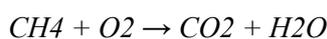 CO2 + H2O

*The challenge is to ensure that the number of atoms of each element is equal on each side.*

**Bing Chat:** *Of course! A chemical reaction occurs when a material undergoes a transformation in which its constitution changes, that is, its atoms rearrange to form new substances. Chemical reactions are represented and interpreted using chemical equations1.*

*For example, when you light a candle, the wax melts and rises up the wick. When the wax hits the candle flame, it burns and produces carbon dioxide and water. This is a simple chemical reaction that produces new substances2. I hope this helps!*

Overall, ChatGPT's response appears to be more effective at addressing the student's difficulty with understanding chemical reactions. The response is more relatable and easier to understand as the use of everyday examples and analogies helps to make the concept more accessible and less abstract for the student. While Bing Chat's response furnishes a basic example to demonstrate the concept of chemical reactions, it lacks the engaging quality present in ChatGPT's one.

**P2:** *wait. that's combustion, isn't it? how do you burn something and get gas and a liquid?*

**ChatGPT:** *Yes, you are correct. Combustion is a burning process, and involves the reaction of a substance with oxygen, usually producing heat and light. In the case of methane combustion (which is a gas), the reaction releases energy in the form of heat and light, while transforming the reactants into new products.*

*Let's look at the methane combustion reaction in more detail:*

CH4 (g) + O2 (g) 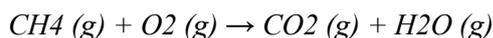 CO2 (g) + H2O (g)

*In this case, the letters (g) indicate that the substances are in the gaseous phase. During combustion, methane (CH4) and oxygen (O2) react together and form carbon dioxide (CO2) and water vapor (H2O), which are also in the gaseous phase due to the high temperatures involved in the reaction.*

*It is important to note that although water is generally associated with the liquid phase, it can also exist as a vapor (gas) at high temperatures, as is the case in this combustion reaction.*

*So, in methane combustion, you start with two gases (methane and oxygen) and produce two new gases (carbon dioxide and water vapour). The transformation of reactants into products is what constitutes a chemical reaction.*

**Bing Chat:** *Yes, you are correct. Burning a candle is an example of combustion. The paraffin is the fuel and the wick is the heat source. When the wick is lit, it heats the paraffin which combines with the oxygen in the air and is burned. Combustion produces carbon dioxide and water, as well as soot (carbon), CO and H2O vapour1. I hope this helps!*

ChatGPT's answer is more comprehensive and directly addresses the student's question, clarifying the misconception about water being only a liquid by mentioning its existence as a vapor at high temperatures. Bing Chat's answer, while giving an example of combustion, does not fully address the student's initial confusion about how combustion can produce a gas and a liquid and lacks the depth provided by ChatGPT. In this case, ChatGPT's answer is of higher quality.

**P2:** *ah. this thing also confuses me. what is the difference between atom, molecule and mole?*

**ChatGPT:** *I understand that these terms can be confusing. Let's clarify each of them:*

*Atom: An atom is the smallest unit of a chemical element that retains its characteristic properties. Atoms are made up of subatomic particles such as protons, neutrons, and electrons. For example, a hydrogen atom (H) is the smallest unit of the element hydrogen.*

*Molecule: A molecule is a group of atoms held together by chemical bonds. Molecules can be composed of atoms of the same element or of atoms of different elements. For example, a water molecule (H2O) consists of two hydrogen atoms (H) bonded to an oxygen atom (O).*

*Mol: A mole is a unit of measurement used in chemistry to express quantities of substances. One mole is defined as the number of elementary entities (atoms, molecules, ions, etc.) in 12 grams of carbon-12, which is approximately 6.022 x 10$^{23}$ entities. This quantity is called Avogadro's number.*

*The idea of the mole is similar to the concept of a dozen, which represents 12 units of something. Just as we can say we have a dozen eggs, we can say we have a mole of atoms or molecules.*

*For example, one mole of water (H2O) contains 6.022 x 10$^{23}$ water molecules. Likewise, one mole of hydrogen atoms (H) contains 6.022 x 10$^{23}$ hydrogen atoms.*

*So, in summary:*

*Atom is the smallest unit of an element.*

*Molecule is a group of atoms bonded together.*

*Mol is a unit of measurement that represents a specific amount of elementary entities (atoms, molecules, ions, etc.).*

*Hope this helps clarify the difference between atom, molecule and mole. If you have any further questions, please feel free to ask!*

***Bing Chat:*** *An atom is the smallest unit of matter that retains the properties of an element1. A molecule is a group of two or more atoms held together by chemical bonds2. A mole is a unit of measurement used in chemistry to express amounts of a chemical substance3. One mole contains Avogadro's number of particles3. I hope this helps!*

ChatGPT's answer thoroughly addresses the student's question and provides examples and an analogy to help the student understand the concept of a mole. Bing Chat's answer is less detailed and less effective in clarifying the concepts for the student. Once again, ChatGPT's answer is of higher quality.

***P2:*** *so can you talk about a mole of beer bottles?*

***ChatGPT:*** *Although the concept of "mole" is used specifically in the context of chemistry to represent a quantity of elementary entities, you could theoretically apply the idea of a "mole" of beer bottles as a way of expressing an enormous quantity of bottles.*

*One mole of beer bottles would equal $6.022 \times 10^{23}$ bottles. However, this amount is extremely large and not practical for everyday use.*

*The idea of the mole was introduced to deal with extremely large quantities of very small entities, such as atoms and molecules, that cannot be counted directly. For larger, more easily counted objects, such as beer bottles, more common and practical units of measurement, such as dozens or hundreds, are more appropriate.*

*So, while it's possible to use the idea of a "mole" to describe any number of objects, this unit is more relevant and useful in the context of chemistry.*

***Bing Chat:*** *Not exactly. The mole is a unit of measurement for expressing the amount of matter in a system. A beer bottle contains many molecules, but it is not possible to tell how many moles of molecules are present in the bottle without knowing the molar mass of the molecule and the mass of the beer bottle. I hope this helps!*

ChatGPT's answer is more relevant and informative, directly addressing the student's question and explaining the practical limitations of using the term "mole" for large, easily counted objects. Bing Chat's answer focuses on the difficulty of determining the number of moles of molecules in a beer bottle, not directly addressing the student's question, and lacking the same level of clarity.

***P2:*** *You mentioned water and water vapor. this is what my professor calls a change of state, right?*

***ChatGPT:*** *Yes, that's right! A change of state, also called a phase transition, is the process by which a substance changes from one physical state to another. There are three common physical* states of matter: solid, liquid, and gas. Changes of state occur when a substance gains or loses energy, usually in the form of heat.

Water is a great example to illustrate state changes, as it can exist in all three physical states under common conditions on Earth:

Solid: Water in the form of ice is a solid. When ice gains heat, it melts and turns into liquid water, undergoing a change of state from solid to liquid.

Liquid: Water in its liquid state is the form we normally know. When liquid water gains heat, it evaporates and turns into water vapor (gas), undergoing a change of state from liquid to gas.

Gaseous: Water in the form of water vapor is a gas. When water vapor loses heat, it condenses and turns into liquid water, undergoing a change of state from gaseous to liquid.

State changes can also occur directly between the solid and gaseous states, without going through the liquid state. This process is called sublimation (solid to gas) or deposition (gas to solid). However, these changes are less common for water under normal conditions on Earth.

Hope this helps to clarify the concept of change of state. If you have more questions or want to explore other examples, let me know!

**Bing Chat:** Yes, exactly! When the candle burns, the wax melts and turns into liquid. Then the liquid is vaporized and turns into wax vapour. The wax vapor is then burned in the candle flame to produce carbon dioxide and water vapour. The change of state occurs when the wax melts and turns into a liquid and when the liquid is vaporized and turns into a wax vapour. I hope this helps!

ChatGPT's answer is more relevant and directly addresses the student's question about water and water vapor, providing a thorough explanation of state changes. Conversely, Bing Chat reverts back to the example of a burning candle, instead of focusing on the water and water vapor example mentioned by the student. The answer, while informative, does not focus on the student's specific example and is less relevant to their inquiry.

**P2:** my teacher said that plasma is the fourth state of matter. but isn't plasma that thing they inject into a vein in the hospital?

**ChatGPT:** I understand the confusion! The word "plasma" has different meanings in different contexts. Let's clarify both uses of the term:

Plasma as the fourth state of matter: In the context of physics and chemistry, plasma is considered the fourth state of matter, along with solid, liquid and gaseous states. Plasma is an ionized state, in which atoms or molecules are energized to the point where they lose electrons, forming positive ions and a cloud of free electrons. This ionization gives the plasma some unique properties, such as the ability to conduct electricity and respond to magnetic fields.

*Plasma is found in some natural phenomena such as lightning and auroras, as well as in man-made devices such as fluorescent lights and controlled nuclear fusion.*

*Blood plasma: In the context of biology and medicine, plasma refers to the liquid component of blood. Blood is made up of blood cells (red blood cells, white blood cells, and platelets) and blood plasma. Blood plasma is a mixture of water, salts, proteins and other dissolved substances. It has many functions, such as transporting nutrients and waste, regulating fluid balance, and helping with blood clotting. When you mention "that stuff they inject into a vein at the hospital," you're probably referring to blood plasma transfusions, which can be used to treat a variety of medical conditions.*

*So, despite sharing the same name, plasma as the fourth state of matter and blood plasma are completely different concepts. The word "plasma" has different meanings in the contexts of physics and medicine.*

**Bing Chat:** *Your teacher is correct! Plasma is the fourth state of matter. It is an ionized gas that contains free electrons and positive and negative ions1. Plasma is formed when a substance in the gaseous state is heated to such a high temperature that the molecular thermal agitation exceeds the binding energy that keeps the electrons in orbit around the nucleus of the atom2. Plasma is found in many places in the universe, including stars such as the Sun3. I hope this helps!*

ChatGPT's answer is more comprehensive and directly addresses the student's confusion about the two different meanings of the word "plasma." It provides a clear distinction between plasma as the fourth state of matter and blood plasma in the medical context. Bing Chat's answer focuses on the physics and chemistry context of plasma as the fourth state of matter and is less relevant to the student's specific question compared to ChatGPT's answer.

**P2:** *another thing that complicates me is these things that you can't see and have to believe exist, like atom, molecule, ion*

**ChatGPT:** *It's true that atoms, molecules and ions are too small to be seen with the naked eye, and that can make them difficult to understand. However, these fundamental chemistry concepts have been developed based on a wide variety of experimental and theoretical evidence over time.*

*Although you cannot directly see an atom or molecule, many of its properties can be observed and measured through experiments and advanced analysis techniques. Here are some ways we can infer the existence of these entities and study their properties:*

*Spectroscopy: This is a technique that involves the interaction of substances with light or other forms of electromagnetic radiation. Spectroscopy provides detailed information about the*

*structure and composition of atoms and molecules, such as binding energies and distances between atoms.*

*X-ray diffraction: X-ray diffraction is a technique used to determine the three-dimensional structure of molecules, especially crystalline molecules. The X-rays are scattered by the atoms in the molecule, and the resulting diffraction pattern can be used to deduce the position of the atoms and the geometry of the molecule.*

*Electron Microscopes: Electron microscopes use beams of electrons instead of light to view objects at extremely small scales. They allow direct observation of nanoscale structures such as crystals and even individual molecules in some cases.*

*Chemical reactions: The behavior of substances in chemical reactions helps us to understand the properties of the atoms, molecules and ions involved. For example, the law of conservation of mass states that the total mass of the reactants equals the total mass of the products, which suggests that atoms are rearranged during the reaction.*

*Believing in the existence of atoms, molecules and ions is a matter of accepting the scientific evidence and theories that have been developed to explain a wide variety of chemical phenomena. As you study chemistry and learn about the different evidence and experiments that support these theories, you will begin to better understand the importance of these concepts and how they relate to the world around us.*

**Bing Chat:** *I understand that it can be hard to believe in things we can't see. The atom is the basic unit of matter, that is, the smallest portion into which an element can be divided without losing its chemical properties. Atoms are formed by a nucleus composed of particles of protons and neutrons and by electrons that orbit the nucleus, forming the electrosphere1. Although we cannot see atoms directly, we can observe their indirect effects. For example, we can see the colors of a rainbow because white light is split into different wavelengths when it passes through raindrops. Similarly, we can see the colors of a flame because the electrons in atoms are excited and emit light when they return to their ground state1. I hope this helps!*

ChatGPT's answer empathizes with the student's difficulty and is more comprehensive and informative, addressing the student's concern by explaining a variety of techniques and experiments that have been developed to study atoms, molecules, and ions. The response emphasizes that believing in their existence is based on the scientific evidence and theories that explain various chemical phenomena. Bing Chat's answer, while acknowledging the student's difficulty and being informative, does not provide as much detail or as many examples.

**P2:** *the teacher said that spectroscopy has to do with fireworks*

**ChatGPT:** *Yes, spectroscopy is related to fireworks in an indirect way. Fireworks are a visual demonstration of the spectroscopic properties of the chemical elements present in them. The*

*bright, vivid colors you see in fireworks are the result of the excitation and emission of light by atoms or ions of these elements.*

*When fireworks are launched, a chemical reaction takes place that releases a large amount of energy. This energy is transferred to the electrons of the atoms of the compounds present in the firework. Electrons, when excited, jump to higher energy levels and then return to their original energy levels. During this process, energy is released in the form of light. The color of the emitted light depends on the amount of energy released, which is characteristic of each element. Spectroscopy is the study of the interaction of light (or other forms of electromagnetic radiation) with matter. In this context, spectroscopy allows us to identify the elements present in fireworks based on the colors they emit. For example:*

*Sodium (Na): Yellow*

*Barium (Ba): Green*

*Strontium (Sr) and Lithium (Li): Red*

*Copper (Cu): Blue*

*Calcium (Ca): Orange*

*So fireworks are a practical and visual way to see the application of spectroscopy in everyday life. They also illustrate how chemistry and physics underlie the phenomena we experience and appreciate.*

**Bing Chat:** *Spectroscopy is a technique used to analyze the light emitted or absorbed by an object. The light is divided into its component colors (a spectrum) and the intensity of each color is measured. Each chemical element has a unique spectrum of colors that it emits or absorbs when it is heated or excited. This means that we can use spectroscopy to identify chemical elements in unknown samples1. In the case of fireworks, chemical elements are heated and excited to produce the brilliant colors we see in the sky. Each color is produced by a different chemical element and has a unique spectrum of colors2. I hope this helps!*

Both ChatGPT and Bing Chat provide informative answers that connect spectroscopy to fireworks, explaining how the colors produced in fireworks can be traced back to the unique spectra of different chemical elements. However, ChatGPT's answer gives more specific examples of elements and their corresponding colors, providing a clearer connection between the two concepts.

**P2:** *The teacher said the copper color is blue. but copper is reddish*

**ChatGPT:** *You are correct in noting that copper in its metallic form has a reddish color. However, the color of a pure metal and the color of light it emits in a firework are not necessarily related.*

*In fireworks, colors are created by the energy released when electrons in atoms or ions are excited and then return to their original energy levels. The specific colors you see are*

*determined by the energy differences between these levels, which are characteristic of each element.*

*In the case of copper, when its electrons are excited and return to their original energy levels, the energy released in the form of light is in the blue range of the visible spectrum. To obtain the blue color in fireworks, copper compounds such as copper chloride (CuCl) or copper oxide (CuO) are usually used.*

*Therefore, although metallic copper is reddish, the light it emits when its electrons are excited and return to their original energy levels is blue. This highlights the distinction between the color of the metal itself and the color of light it emits when involved in a chemical reaction, as in fireworks.*

**Bing Chat:** *Copper is used in fireworks to produce the color blue. When copper is heated, it emits a bluish light1. I hope this helps!*

ChatGPT's answer offers a more comprehensive explanation for why copper emits blue light in fireworks, despite its reddish color as a metal. The response delves into the details of energy levels and electron excitation, while Bing Chat's answer simply states the observed phenomenon without providing the underlying reasoning. Nevertheless, both failed in explaining why metallic copper is reddish.

***P2:*** *I also get confused with covalent bonding, ionic bonds and hydrogen bonds*

***ChatGPT:*** *It's important to understand the differences between these three types of bonds, as they play key roles in the chemistry and properties of compounds. Let's take a look at each type of binding:*

*Covalent Bonding: Covalent bonding occurs when two atoms share one or more pairs of valence electrons. This sharing of electrons allows both atoms to reach a stable electron configuration, usually with an octet of electrons in their valence shells. Covalent bonds are common between atoms of non-metallic elements.*

*For example, in the water molecule ($H_2O$), each hydrogen atom shares a pair of electrons with the oxygen atom, forming two covalent bonds.*

*Ionic Bonding: Ionic bonding occurs when an atom donates one or more electrons to another atom, resulting in the formation of oppositely charged ions that are attracted to each other due to electrostatic force. Ionic bonds are common between atoms of metallic and non-metallic elements.*

*For example, in sodium chloride (NaCl), the sodium atom donates an electron to the chlorine atom, forming the sodium ion ($Na^+$) and the chloride ion ($Cl^-$). These ions are attracted to each other due to their opposite charges, forming an ionic bond.*

*Hydrogen Bonding: Hydrogen bonding is a special type of intermolecular interaction, not a true chemical bond like covalent and ionic bonds. Hydrogen bonding occurs when a hydrogen atom*

*covalently bonded to an electronegative atom (such as oxygen, nitrogen, or fluorine) is attracted to another nearby electronegative atom.*

*For example, in water molecules, the hydrogen in a water molecule is covalently bonded to oxygen, but is also attracted to the oxygen in other nearby water molecules, forming hydrogen bonds. These hydrogen bonds are responsible for many of water's unique properties, such as its high boiling point and its ability to dissolve a wide variety of substances.*

*Understanding the differences between covalent, ionic, and hydrogen bonds is critical to understanding the chemistry and properties of compounds, as well as their interactions and behavior in different situations.*

**Bing Chat:** *Covalent bonds occur when two atoms share electrons to form a molecule1. Ionic bonds occur when an atom donates electrons to another atom to form an ionic compound2. Hydrogen bonds are a special form of covalent bonding that occurs between a hydrogen atom and an electronegative atom (such as oxygen or nitrogen) in another molecule3. I hope this helps!*

Both ChatGPT and Bing Chat provide informative answers explaining each bond type. ChatGPT's answer is more comprehensive and informative, providing clear examples to illustrate the differences between the bond types. The response also highlights the significance of understanding these bonds in the context of chemistry and the properties of compounds. Bing Chat's answer is more concise, lacking the depth and clarity of ChatGPT's response.

**P2:** *so ionic bonding is when the atom gives the electron once and for all, in the covalent bond, it only lends and in the hydrogen bond it's like a bid, like the electron is with an atom but gives the atom a wink neighbor?*

**ChatGPT:** *This is a simplified and humorous way to think about the differences between ionic, covalent, and hydrogen bonds, but it captures the essence of each type of interaction:*

*Ionic bond: It is as if the atom "gave" its electron once and for all to another atom, resulting in the formation of ions with opposite charges that are attracted to each other. This happens mainly between metallic and non-metallic elements.*

*Covalent bond: It is as if the atoms "borrowed" electrons to each other, sharing pairs of valence electrons. This mainly occurs between atoms of non-metallic elements and allows both atoms to reach a stable electronic configuration.*

*Hydrogen bonding: It is like a "flirting" between a hydrogen atom covalently bonded to an electronegative atom and another electronegative atom nearby. This attraction is weaker than a covalent or ionic bond and is responsible for many of the specific properties of compounds such as water.*

*It is important to remember that this description is an oversimplification and, in reality, interactions between atoms are governed by electrostatic forces and more complex quantum principles. However, this analogy can help you remember and distinguish different types of calls and interactions.*

***Bing Chat:*** *That's right! In ionic bonding, one atom donates electrons to another atom to form an ionic compound. In covalent bonding, the two atoms share electrons to form a molecule. And in hydrogen bonding, a hydrogen atom bonds with an electronegative atom (such as oxygen or nitrogen) in another molecule. I hope this helps!*

ChatGPT acknowledges the simplified and humorous interpretation, agreeing that it captures the essence of each type of interaction. ChatGPT's answer is more comprehensive and clarifies that this is an oversimplification but still useful for understanding the differences between bond types. Bing Chat's answer is shorter and does not emphasise the oversimplified nature of the interpretation or its usefulness.

**DISCUSSIONS**

The overall success of this educational journey, with GenAIbots serving as agents-to-think-with, highlights several powerful pedagogical strategies that can facilitate a profound comprehension of the involved concepts. A significant part of the process involved positive reinforcement, as GenAIbots acknowledged students' observations as enjoyable, thereby validating their experiences and stimulating their curiosity. Furthermore, a Socratic-like questioning approach was adopted, prompting students to think critically and actively engage in their learning rather than merely providing them with answers. GenAIbots also capitalised on the student's existing knowledge by inquiring about their understanding of topics like weight and gravity, thereby establishing a meaningful bridge between new and previously learned information.

The interactive nature of this learning experience allowed students to expand their existing knowledge incrementally. GenAIbots offered relevant explanations, examples, and analogies to support their learning, adapting responses to the students' questions and confusions, providing customised explanations, and addressing misconceptions.

This study underscores the potential of AI-driven tools like GenAIbot in promoting active learning, personalised instruction, and conceptual understanding across various domains, including physics. It also emphasises the value of AI in enhancing educational experiences. It concludes that AI-driven chatbots, such as ChatGPT and Bing Chat, can act as innovative platforms for exploring a wide range of scenarios often challenging for students.

Nevertheless, our analysis reveals that ChatGPT consistently provided more comprehensive, detailed, and accurate answers than Bing Chat, which tended to give shorter and less informative responses. ChatGPT was also more adept at addressing the nuances and subtleties of the questions, recognising the context, and engaging with the user's unique perspective or interpretation. On the other hand, Bing Chat often failed to address these nuances and provided less contextual information. Overall, ChatGPT demonstrated a superior understanding of the subject matter and a more remarkable ability to convey complex scientific concepts in an accessible manner.

As the study suggests, using well-crafted sentences or prompt crafting can mitigate the risk of misinterpretation and support the assessment of students' alternative conceptions within the context of scientific knowledge. Crafting prompts that elicit the desired response becomes crucial to the iterative reflections and interactions when using GenAIbots.

The study also highlighted the need for comprehensive educator training before integrating these tools into classrooms. Adequate training empowers teachers to guide students effectively in articulating their thoughts and constructing solid arguments. This preparation allows educators to assess GenAIbots' performance when faced with compelling counterexamples, fostering productive discussions among colleagues.

By leveraging the potential of generative AI chatbots as educational tools, teachers can design more engaging and personalised learning experiences. This approach fosters a deeper understanding of complex concepts and promotes critical thinking across various subjects.

**CONCLUSIONS**

In this paper, we analysed the performance of two AI language models, ChatGPT and Bing Chat, responding to a series of questions related to chemistry, physics, and scientific concepts. The questions covered topics such as plasma, atoms, molecules, ions, spectroscopy, fireworks, and various types of chemical bonds. Notably, ChatGPT consistently outperformed Bing Chat, providing more comprehensive, detailed, and accurate responses and skillfully addressing nuances and context.

Nevertheless, our findings suggest that GenAIbots, including ChatGPT and Bing Chat, are robust agents-to-think-with in STEM education. Their benefits include interactivity, customisation, multidisciplinary knowledge, critical thinking, problem-solving, creativity, and collaboration. GenAIbots foster dynamic conversations and personalised content and offer a wealth of knowledge across subjects.

Interaction with GenAIbots aids in refining critical thinking, problem-solving, and creativity. They facilitate discussions, debates, and collaborative projects in group settings, thereby deepening topic understanding.

Aligned with Seymour Papert's concept, GenAIbots are exceptional agents-to-think-with, fostering an engaging and inclusive learning environment. They encourage creativity, collaboration, exploration, critical thinking, and problem-solving.

Incorporating GenAIbots into STEM education supports the constructionist learning theory, emphasising experiential, hands-on learning. Utilising GenAIbots, students can refine their problem-solving, critical thinking, and creative skills in an engaging learning environment, preparing them for success in STEM fields.

**Limitations of the study and possible future studies**

Despite its inherent limitations, including its single case design and the potential for bias, this study yielded valuable insights from extensive interactions between students and GenAIchats, specifically ChatGPT and Bing Chat, the two most widely used at the time. The exploratory depth of the study uncovered hidden potential within these systems, even amid concerns about generalizability.

Future research could include:

- extending to emerging chatbots, like the emerging Bard.
- focusing on refining the crafting of prompts.
- investigating whether a casual or more formal communication tone from GenAIbots leads to enhanced learning and how offering this tonal choice to students might customise their learning experience further.
- examining if a GenAIbot impersonating a peer or an expert renowned for science communication, such as Neil de Grasse-Tyson, Alex Filippenko, Andrew Fraknoi, or Fay Dowker, leads to improved learning and how offering this choice of 'persona' to students might further tailor their learning experience.

When implementing GenAIbots in STEM learning, it is crucial to carefully evaluate benefits and drawbacks, ensuring accurate information delivery and considering the implications of reduced human interaction. These concerns may be alleviated by integrating GenAIbots with other educational tools or activities promoting collaborative dialogue among learners.